\documentclass[%
 reprint,superscriptaddress,
 amsmath,amssymb,
 aps,
]{revtex4-2}
\usepackage{epsfig}
\usepackage{upgreek}
\usepackage{amsmath}
\usepackage{times}
\usepackage{siunitx}
\usepackage{braket}
\usepackage{booktabs}
\usepackage{chemist}
\usepackage{mathtools}
\usepackage{float}
\usepackage{comment}
\usepackage{appendix}
\usepackage{graphicx}% Include figure files
\usepackage{dcolumn}% Align table columns on decimal point
\usepackage{bm}% bold math
\usepackage{natbib}
\usepackage{textcomp}
\usepackage{xspace}
\usepackage{xcolor}
\usepackage{booktabs}
\usepackage{multirow}

\DeclareSIUnit{\sample}{Sa}
\newcommand{\ReMo}[1]{{\color[rgb]{0,0,0}#1}}

\usepackage{titlesec}
\titlespacing\section{0pt}{12pt plus 6pt minus 2pt}{8pt plus 2pt minus 0pt}
\titlespacing\subsection{0pt}{12pt plus 6pt minus 2pt}{8pt plus 2pt minus 0pt}
\titlespacing\subsubsection{0pt}{12pt plus 6pt minus 2pt}{8pt plus 2pt minus 0pt}

\let\OldTexttrademark\texttrademark
\renewcommand{\texttrademark}{\OldTexttrademark\xspace}%

\begin{document}

\setlength{\abovedisplayskip}{3pt}

\preprint{APS/123-QED}

\title{Microwave Package Design for Superconducting Quantum Processors}

\author{Sihao Huang}
\thanks{These two authors contributed equally}
\affiliation{Department of Physics, Massachusetts Institute of Technology, Cambridge, MA 02139, USA}
\affiliation{Department of Electrical Engineering and Computer Science, Massachusetts Institute of Technology, Cambridge, MA 02139, USA}
\affiliation{Research Laboratory of Electronics, Massachusetts Institute of Technology, Cambridge, MA 02139, USA}

\author{Benjamin Lienhard}
\thanks{These two authors contributed equally}
\affiliation{Department of Electrical Engineering and Computer Science, Massachusetts Institute of Technology, Cambridge, MA 02139, USA}
\affiliation{Research Laboratory of Electronics, Massachusetts Institute of Technology, Cambridge, MA 02139, USA}

\author{Greg Calusine}
\affiliation{MIT Lincoln Laboratory, Lexington, MA 02421, USA}

\author{Antti Veps{\"a}l{\"a}inen}
\affiliation{Research Laboratory of Electronics, Massachusetts Institute of Technology, Cambridge, MA 02139, USA}

\author{Jochen Braum{\"u}ller}
\affiliation{Research Laboratory of Electronics, Massachusetts Institute of Technology, Cambridge, MA 02139, USA}

\author{David K. Kim}
\affiliation{MIT Lincoln Laboratory, Lexington, MA 02421, USA}

\author{Alexander J. Melville}
\affiliation{MIT Lincoln Laboratory, Lexington, MA 02421, USA}

\author{Bethany M. Niedzielski}
\affiliation{MIT Lincoln Laboratory, Lexington, MA 02421, USA}

\author{Jonilyn L. Yoder}
\affiliation{MIT Lincoln Laboratory, Lexington, MA 02421, USA}

\author{Bharath Kannan}
\affiliation{Department of Electrical Engineering and Computer Science, Massachusetts Institute of Technology, Cambridge, MA 02139, USA}
\affiliation{Research Laboratory of Electronics, Massachusetts Institute of Technology, Cambridge, MA 02139, USA}

\author{Terry P. Orlando}
\affiliation{Department of Electrical Engineering and Computer Science, Massachusetts Institute of Technology, Cambridge, MA 02139, USA}
\affiliation{Research Laboratory of Electronics, Massachusetts Institute of Technology, Cambridge, MA 02139, USA}

\author{Simon Gustavsson}
\affiliation{Research Laboratory of Electronics, Massachusetts Institute of Technology, Cambridge, MA 02139, USA}
% may require new affiliation (keysight, labber)

\author{William D. Oliver}
\affiliation{Department of Electrical Engineering and Computer Science, Massachusetts Institute of Technology, Cambridge, MA 02139, USA}
\affiliation{Research Laboratory of Electronics, Massachusetts Institute of Technology, Cambridge, MA 02139, USA}
\affiliation{MIT Lincoln Laboratory, Lexington, MA 02421, USA}

\begin{abstract}

Solid-state qubits with transition frequencies in the microwave regime, such as superconducting qubits, are at the forefront of quantum information processing. However, high-fidelity, simultaneous control of superconducting qubits at even a moderate scale remains a challenge, partly due to the complexities of packaging these devices. Here, we present an approach to microwave package design focusing on material choices, signal line engineering, and spurious mode suppression. We describe design guidelines validated using simulations and measurements used to develop a 24-port microwave package. Analyzing the qubit environment reveals no spurious modes up to \SI{11}{\giga\hertz}. The material and geometric design choices enable the package to support qubits with lifetimes exceeding \SI{350}{\micro\second}. The microwave package design guidelines presented here address many issues relevant for near-term quantum processors. 

\end{abstract}

\maketitle

Over the past two decades, superconducting qubits have emerged as one of the leading candidates for building large-scale quantum systems~\cite{krantz_kjaergaard_yan_orlando_gustavsson_oliver_2019, Nori2017_review}. Today, superconducting qubits routinely reach coherence times in the range of \SI{100}{\micro\second}~\cite{Jin2015_PRL_therm}, achieve gate times of a few tens of nanoseconds~\cite{Arute2019_supremacy}, and have exhibited single- and two-qubit gates with fidelities exceeding the threshold for the most lenient quantum error correction codes~\cite{state_of_play}. The characteristics of these qubits and their surrounding circuit elements for control and readout can be engineered with high precision using well established fabrication techniques~\cite{Schoelkopf2008_Nature_Wire} leveraging standard tools developed for the semiconductor industry. 

Despite the rapid progress towards building practical quantum processors using superconducting qubits~\cite{Arute2019_supremacy}, the susceptibility of these artificial atoms to noise remains a significant engineering challenge to system scaling. Microwave packaging is a part of this challenge. In particular, package designs must support increasing qubit numbers while also preserving qubit coherence and high-fidelity quantum operations. 

Here, we present an approach to multi-port packaging and discuss the relevant microwave design principles. We designed a 24-port package and validated it using transmon qubits---one of the most commonly employed superconducting qubit modalities. We performed detailed characterizations of the package materials, signal transmission, and mode profile. We experimentally conducted a ``hidden-mode'' survey~\cite{HiddenModeExperiment} of the device package using four transmon qubits with lifetimes reaching \SI{120}{\micro\second}. The considered loss channels enable the presented package to support transmon qubits with lifetimes in excess of \SI{350}{\micro\second}. While the qubits employed in our characterization are not lifetime-limited by the package, the limit is within the same order of magnitude as the lifetime of state-of-the-art transmons. This underscores the importance of further improvements in microwave engineering to minimize qubit energy loss channels. These engineering principles provide tools for the development of improved packaging for near-term quantum processors~\cite{Preskill2018_supremacy}. 

We introduce the \ReMo{design considerations} for qubit packaging in Sec.~\ref{sec:req} and consider materials, signal routing, and modes in Sec.~\ref{sec:mat}-\ref{sec:modes} with discussions on characterization and performance validation. Finally, we discuss how the presented microwave design principles generalize and conclude in Sec.~\ref{sec:con}.

\section{\label{sec:req}Superconducting Qubit Microwave Environment}

\begin{figure}
\includegraphics[width=0.47\textwidth]{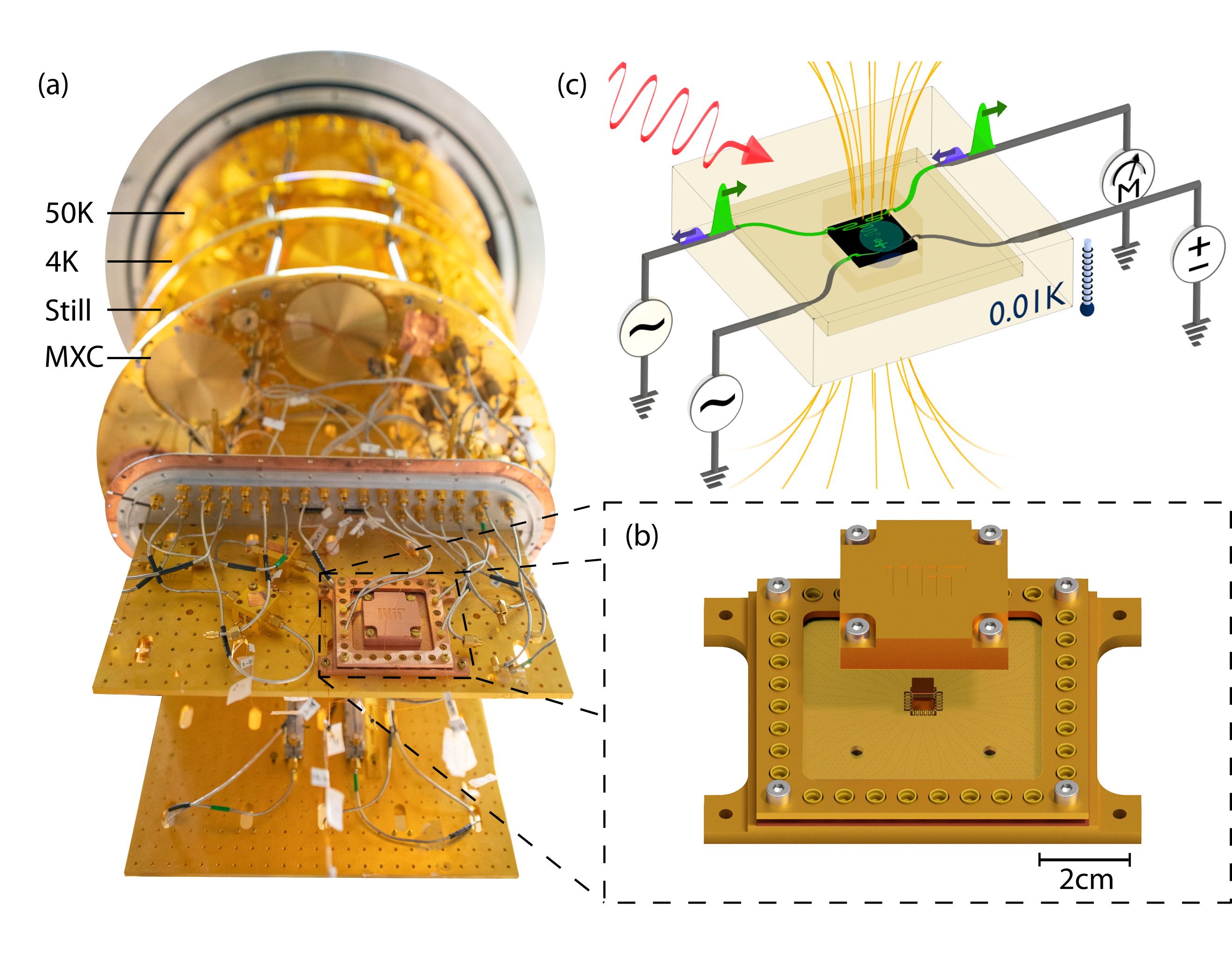}
\caption{\label{fig:overview}Package environment, layout, and requirements. (a) Dilution refrigerator with multiple temperature stages holding the qubit chip enclosed in a microwave package. The microwave package interfaced with through microwave lines is mounted on a cold finger in the mixing chamber reaching a base temperature of approximately \SI{10}{\milli\kelvin}. (b) The microwave package consists of a metal enclosing, microwave connectors, an interposer for signal fanout, and a microwave cavity in the center surrounding the quantum chip. (c) The purpose of a microwave package is to shield the enclosed qubit chip from external radiation (red oscillating arrow) and stray magnetic fields (yellow lines) while providing impedance-matched (\ReMo{transmitted green pulse and} reflected blue \ReMo{pulse at the input and} output), low-crosstalk communication channels (crosstalk in green at input), and a thermal link to the dilution refrigerator.}
\end{figure}

The material composition and thermal environment of a qubit define its feasible operational frequency range. Superconducting qubits are generally designed to have transition frequencies between \SI{2}{\giga\hertz} and \SI{10}{\giga\hertz}~\cite{Oliver2013_review}. The qubits are shielded and cooled in a \textsuperscript{3}He-\textsuperscript{4}He dilution refrigerator to minimize thermal excitations. Today's commercially available dilution refrigerators reach a base temperature around \SI{10}{\milli\kelvin}, \ReMo{well below the temperature corresponding to the transition frequency of the qubit (which is around \SI{240}{\milli\kelvin} for a \SI{5}{\giga\hertz} qubit) and the critical temperature of superconductors used for qubit design (e.g., \SI{1.2}{\kelvin} for aluminum).} The device package is mounted on the mixing chamber plate, the coldest stage in the refrigerator, shown in Fig.~\ref{fig:overview}(a). We designed a superconducting qubit package [Fig.~\ref{fig:overview}(b)] consisting of a copper casing, a multilayer interposer to perform signal fanout~\cite{Bronn2018_package}, a shielding cavity in the package center, and a set of microwave connectors or integrated flex-cables~\cite{RigettiFlex}. 

The purpose of a microwave package is threefold, depicted in Fig.~\ref{fig:overview}(c). First, the package casing suppresses the coupling of the qubits to decoherence channels external to the package, such as environmental electromagnetic noise. Next, the package accommodates qubit control channels to and from the enclosed quantum processor. Finally, the package sinks excess and latent thermal energy due to qubit control and readout operations. However, a microwave package---the immediate qubit environment---can induce losses of its own if not carefully engineered. Designing a microwave package requires a balance between the suppression of external and package-induced loss channels. 

The different loss channel contributions can be indirectly evaluated and characterized through qubit coherence, which is affected by interactions with its environment~\cite{Nielsen_Chuang_2011}. These interactions lead to energy decay and dephasing. Qubit coherence generally improves as the qubit is decoupled from the environment, but it becomes more difficult to control and read out. Ideally, the qubit would couple exclusively to the control and readout environment.

In thermal equilibrium, the excitation probability of a qubit can be expressed with the Boltzmann factor \(\exp{(-h \nu/k_{\rm B} T)}\) (\(\nu\): qubit transition frequency, \(T\): qubit temperature, \(h\): Planck constant, \(k_{\rm B}\): Boltzmann constant). However, thermal equilibrium with \SI{10}{\milli\kelvin} is generally not reached due to the influx of thermal photons from higher temperature stages of the refrigerator via the signal lines. With state of the art attenuation and filtering~\cite{Serniak2018_T}, superconducting qubits have achieved effective temperatures of \SI{35}{\milli\kelvin}, corresponding to an excited state thermal population of \(0.1\%\) at \SI{5}{\giga\hertz}~\cite{Jin2015_PRL_therm}. As a consequence, the qubit energy exchange with its environment is approximately a unidirectional energy decay from the qubit to the environment. The rate at which the qubit loses its energy is referred to as the longitudinal relaxation rate \(\Gamma_{1} \triangleq T_1^{-1}\), with \(T_1\) the energy relaxation time. Energy decay is mediated by various loss channels, such as \ReMo{quasiparticles, vortices,} surface dielectric dissipation, conductivity losses, or dissipation into spurious package modes near the qubit transition frequency. The quality factor \( Q_i\) expresses the inverse `lossiness' of an individual loss channel \(i\). The participation ratio \(p_{i}\), a unitless factor, associates each loss channel with a normalized interaction strength between itself and the qubit~\cite{Calusine2018_loss, Woods2018_loss} so that the participation ratios of all loss channels sum to \(1\). Modeling the qubit as a harmonic oscillator---a reasonable approximation for weakly anharmonic qubits like the transmon~\cite{Koch2007_Transmon} and capacitively shunted flux qubits~\cite{Fei2016_FQ}---the energy exchange rate can then be expressed as \(\Gamma_{1} = 2\pi\nu/Q = 2\pi\nu \sum_{i} p_{i}/Q_{i}\). 

The transition frequency between the ground and excited state of the qubit can be affected by its electromagnetic (EM) environment. Fluctuating EM fields detuned from the qubit transition frequency---if coupled to the qubit---can induce qubit energy-level shifts that cause a change in the phase accumulation rate, resulting in pure dephasing \(\Gamma_{\upphi}\) of a qubit superposition state. In addition to pure dephasing, energy relaxation is also a phase breaking process. The dynamics of two-level systems weakly coupled to noise sources can be described using the Bloch-Redfield formalism~\cite{Bloch1957_PR_BRF, Redfield1957_IBM_BRF}, where the combination of pure dephasing \(\Gamma_{\upphi}\) and the process of energy relaxation is given by the transverse relaxation rate \(\Gamma_2 \triangleq 1/T_2^{*} = \Gamma_1/2 + \Gamma_{\upphi}\).

\section{\label{sec:mat}Microwave Package Materials}

Material-dependent losses can be of magnetic (\(1/Q_{m}\)), conductive (\(1/Q_{c}\)), or dielectric (\(1/Q_{d}\)) origin~\cite{PozarLoss}. Energy loss channels couple to the qubit through its electric or magnetic dipole moment. For transmon qubits, the electric dipole moment presently dominates~\cite{Koch2007_Transmon}. Qubits are fabricated using high-\(Q\) materials and substrates to reduce loss. In addition the device geometry is designed to reduce the electric field density in lossy regions, such as surfaces and interfaces~\cite{Lienhard2019_package}. 

Many qubit architectures---in particular those with a tunable transition frequency---are sensitive to magnetic fields. Consequently, magnetic metals or materials with magnetic compounds are generally avoided. To shield qubits from magnetic field fluctuations, materials with high magnetic permeability such as mu-metal can be used either as part of the the dilution refrigerator infrastructure or on the package casing. 
%The magnetic losses scale linearly with magnetic permeability of the material. 
An alternative approach is to incorporate type-I superconductors such as aluminum, tin, or lead, in the package body. Once such a material turns superconducting, it expels the magnetic field from its core due to the Meissner effect, so long as the magnetic field does not exceed a specific material- and temperature-dependent threshold. %Superconductors, typically aluminum, are widely-used to fabricate the package body for this reason.

Conductivity losses arise when the electric field of the qubit induces a current in nearby normal conductors with a finite conductivity. The loss depends on the conductivity \(\sigma\) of the material and scales as \(1/Q_{c} \propto 1/\sqrt{\sigma}\). Losses also arise due to atomic defects in bulk dielectrics hosted on those interfaces absorbing EM energy. Dielectric losses are proportional to the imaginary dielectric coefficient \(\operatorname{Im}(\epsilon)\) of the material \(1/Q_{d} \propto \operatorname{Im}(\epsilon)\).

Commonly employed package materials include superconducting aluminum, copper, and gold-plated copper. Superconducting aluminum forms a thin oxide layer of approximately \SI{2}{\nano\meter}~\cite{Quintana2014_mat}, inducing some dielectric losses while keeping the conductivity losses at a minimum. Similar to aluminum, copper forms an oxide layer~\cite{Fredj2011_Cu} leading to dielectric as well as conductivity loss due to its non-zero resistance~\cite{Campbell1974_Cu}. Gold-plating limits the oxide formation at the cost of an increase in conductivity losses by up to one order of magnitude~\cite{Lide1996_Au}.

The device package demonstrated in this work, shown in Fig.~\ref{fig:overview}(b), is composed of a base and lid, both milled from oxygen free high conductivity (OFHC) copper. To increase the package fundamental mode frequency and suppress material-induced losses, the qubit chip is suspended by at least \SI{3}{\milli\meter} to form a cavity above and below it~\cite{Lienhard2019_package}. A layer of aluminum with a target thickness of \SI{500}{\nano\meter} is evaporated on the lid center cavity surface to reduce conductivity losses. 

Full-wave EM simulations (COMSOL Multiphysics\textregistered) indicate the layer of aluminum on the center cavity surface to reduce the material-induced loss channels by three orders of magnitude. As such, the material-induced losses of the presented package are negligible, enabling it to support qubits with lifetimes up to seconds, see Tab.~\ref{tab:mat}. 

\begin{table}[b]
\center
\setlength{\tabcolsep}{5pt}
\begin{tabular*}{0.48\textwidth}{lccc}
\toprule
Center Cavity & Conductivity Loss &  Dielectric Loss & $T_1$-limit \\ 
& $1/Q_c$ & $1/Q_d$ & (\si{\second}) \\ 
\midrule
Bare Cu             & $\num{2e-9}$      & $\num{1e-12}$     & $0.020$   \\
Al-evaporated Cu    & $-$               & $\num{5e-12}$     & $5.830$   \\
Au-plated Cu        & $\num{5e-9}$      & $-$               & $0.006$   \\ \bottomrule
\end{tabular*}
\caption{\label{tab:mat}Comparison of qubit losses induced by different casing materials for the presented package geometry. The values are obtained with EM simulations of a \(\nu=\SI{5}{\giga\hertz}\) transmon qubit. The qubit is located in the corner of a $\SI{5}{\milli\meter}\times \SI{5}{\milli\meter}$ chip to estimate the maximum impact of package material induced losses. The $T_1$-limit is estimated as \(1/(1/Q_c+1/Q_d)\cdot 1/(2\pi\nu)\). We assumed the following material dependent characteristics: Copper (Cu), conductivity of \SI{5e10}{\siemens\meter}~\cite{Campbell1974_Cu} and an oxide thickness of \SI{10}{\nano\meter}~\cite{Fredj2011_Cu}; Aluminum (Al), conductivity of infinity and an oxide thickness of \SI{2}{\nano\meter}~\cite{Quintana2014_mat}; Gold (Au), conductivity of \SI{5e9}{\siemens\meter}~\cite{Lide1996_Au} and no oxide layer.}
\end{table}

A device package needs to provide an efficient thermal link to the dilution refrigerator to ensure the qubits reach and remain close to the base temperature. The accumulated thermal energy due to all sources, including the material loss channels, should not significantly heat the qubits. Furthermore, as the number of qubits increases, the efficiency of the thermal link has to increase accordingly. Heat is removed by electrical connections to the wiring and attenuation at cryogenic temperatures---where \ReMo{non-equilibrium} electrons can thermalize---and by mechanical (phonon) connections to the package.

The loss channel contributions of the material have to be balanced with its thermal conductivity. The thermal conductivity of copper decreases linearly with temperature and reaches a value of approximately \SI{0.5}{\watt\per\centi\meter\per\kelvin} at \SI{20}{\milli\kelvin}~\cite{Dupre1964_Cu}. On the other hand, the thermal conductivity of superconducting aluminum decreases exponentially faster at similar temperatures~\cite{Sharma1967_Al} and can be estimated to be around \SI{0.025}{\watt\per\centi\meter\per\kelvin} at \SI{20}{\milli\kelvin}. However, at cryogenic temperatures, heat flow from the chip is almost entirely dominated by Kapitza boundary resistance interfaces~\cite{Swartz1989_Kapitza}.

In the presented package, the qubit chip is kept in place with \ReMo{aluminum} wirebonds and pressed down on copper posts located in the corners of the qubit chip. A stronger thermal link could potentially be formed using conductive adhesives such as silver paste or polymeric adhesives. However, we avoided the use of adhesives as their conductivity is typically more than three orders of magnitude smaller compared to copper resulting in a measurable increase in conductivity losses~\cite{Goetz2016_loss}.

\section{\label{sec:crs}Control and Readout Signals}
% two primary focii: crosstalk & impedance matching

\begin{figure*}
\includegraphics[width=1\textwidth]{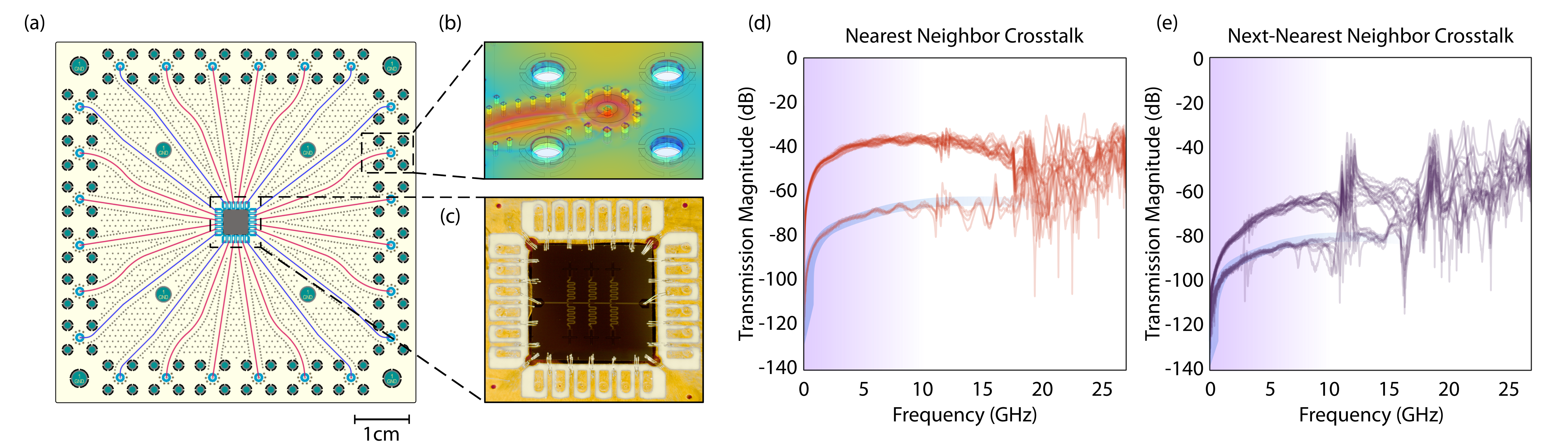}
\caption{\label{fig:interposer}Interposer design. (a) Interposer layout for a 24-line package fabricated out of a three-layer Rogers 4350\texttrademark controlled impedance glass-reinforced ceramic laminate. Stripline-based waveguides with dense via shielding are utilized to reduce signal crosstalk. (b) EM simulation (3D finite element simulation software COMSOL Multiphysics\textregistered) of a via transition used for \ReMo{subminiature push-on (SMP) type} microwave connectors. The transition is impedance matched (not shown) to minimize reflections and effects on the step response. (c) Picture of a \SI{5}{\milli\meter} by \SI{5}{\milli\meter} qubit chip mount with signal launches in the periphery. Wirebonds are used to provide signal connections and ground the device. (d) and (e) Measured nearest neighbor and second nearest signal crosstalk. A high-port count network analyzer using Keysight M9374A PXIe modules was used to obtain the full scattering matrix. All transmission parameters corresponding to signal crosstalk information are overlaid on the plots. The purple background fading out with increasing frequency indicates the decreasing relevance of modes as their frequency separation to the qubit transition frequency increases. Note that the separate group of traces with greater isolation, as highlighted in (d) and (e) correspond to crosstalk that is reduced at corners of the package, as indicated by the pairs of blue lines in (a).}
\end{figure*}

The signal path of a package introduces a variety of factors, including a distorted step response and insertion loss (Sec.~\ref{subsub:step}) as well as crosstalk (Sec.~\ref{subsub:cross}), which can have a bearing on qubit control. We then describe the engineering considerations for waveguides imprinted within the interposer (Sec.~\ref{sub:int}), and wirebonds (Sec.~\ref{sub:wirebonds}) connected to the qubit chip. 

\subsubsection{\label{subsub:step}Step Response and Insertion Loss}

Good impedance matching leads to lower insertion losses and improved signal integrity, both of which are critical for high-fidelity control and readout. For a linear time-invariant system, the ideal temporal response for a step-like input is instantaneous and step-like as well. There are two related measures: the \textit{rise-time} is the time it takes for the signal to reach a desired amplitude, while the \textit{settling time} is the time that elapses for the signal to stabilize at the output, typically after ringing or repeated oscillations. For one-qubit gates, these distortions can lead to under and over-rotations and reduce gate fidelity. In two-qubit gates, such as the controlled-phase gate, deviations from the carefully shaped flux pulses can lead to leakage away from the computational subspace~\cite{dicarlo_chow_gambetta}. Furthermore, a long settling time will make the action of the gate dependent on the history of pulses applied~\cite{dicarlo_et_al_2020, langford_2017}. While a non-ideal step response can be straightforwardly compensated for by an arbitrary waveform generator using predistortion, longer time-scale distortions require greater memory depth. This motivates improved impedance matching.

The step response is determined by the frequency response of the system. In general, the characteristic impedance of a transmission line is given by $Z = \sqrt{\left(R + i\omega L \right)/\left(G + i \omega C\right)},$ where $R$ is the series resistance of the line, $G$ the parallel conductance of the dielectric, $C$ the parallel capacitance, $L$ the inductance, and $\omega$ is the angular frequency of the signal. Thus, for example, frequency dependence can arise in the case of nonzero resistance, skin effects---which reduce the effective inductance at higher frequencies~\cite{hayt_buck_2001}---and the frequency-dependent nature of the permittivity of the dielectric, which can cause $C$ to vary. 

Decreasing the resistance in the signal lines and return path, either by employing superconductors or high conductivity materials, lowers this frequency-dependence and can effectively reduce the settling and rise times of the system~\cite{Foxen2018}. Furthermore, in the equivalent time-domain picture, poor forward and backward matching cause repeated signal reflections in the system that lead to frequency-dependent standing waves, which, in combination with unmatched reactance in the system, can lead to ringing. As a result, geometric transitions and waveguides within the signal chain must be designed to ensure a consistent impedance.

\subsubsection{\label{subsub:cross}Crosstalk}

Second, crosstalk---the undesired transfer of a signal between separate communication lines---needs to be suppressed in a package. Control signal that leak to nearby qubits can induce unwanted gate operations resulting in a reduction in computational performance. Furthermore, these errors may be frequency-dependent and difficult to compensate for if the qubits add significant non-linearity to the system. Such errors are especially harmful to standard quantum error correction codes, which presume errors that are local and uncorrelated~\cite{sarovar2019detecting}.

Crosstalk decreases as the physical distance between signal lines increases. Interactions between adjacent lines are classified into two categories: near-end crosstalk (NEXT)---measured at the same end as the interfering transmitter---and far-end crosstalk (FEXT)---measured at the opposite end [illustrated in Fig.~\ref{fig:wirebond} (a)]. Coupling between adjacent lines occurs through their mutual capacitance and inductance. Capacitive coupling induces a positive current on both ends of the disturbed line, while inductive coupling leads to a current moving parallel to the instigating line. As a result, the crosstalk can interfere with the desired control signal on both ends of the signal line depending on the ratio of capacitive and inductive couplings.

\subsection{\label{sub:int}Interposer design}

There are three common transmission line designs: coplanar waveguides, microstrips, and striplines~\cite{PozarCavities}. Compared to coplanar waveguides and microstrips, a symmetrical buried stripline is surrounded by a homogeneous dielectric. This environment leads to the same capacitive and inductive coupling to the top and bottom ground plane, cancelling out the forward-propagating current and thereby suppressing far-end crosstalk at the cost of higher fabrication complexity. Furthermore, the top and bottom ground planes shield the fully buried signal line from the far-field environment~\cite{Sage2011_loss}.

The presented interposer is constructed using a three-layer, low-loss Rogers 4350\texttrademark laminate composed of glass-woven hydrocarbon and ceramics (detailed discussion of the stackup can be found in Appendix~\ref{appendix:interposer}). The interposer uses symmetric \ReMo{copper} striplines embedded in the printed circuit board (PCB) dielectric protected by via fences. We employ EM simulations and time-domain modeling to minimize impedance mismatches, illustrated in Fig.~\ref{fig:interposer}(b). The simulations ensure that microwave connector transitions, composed of a grounding cage and a signal via, as well as the wirebond launches, are properly impedance matched~\cite{isidoro-munoz_torres-torres_tlaxcalteco-matus_hernandez-sosa_2017, roh_li_ahn_park_cha_2006, keysight_2017}.

The crosstalk is further suppressed using via fences---rows of metallized holes drilled through the substrate material to shield in-plane EM-field coupling between pairs of signal lines. As opposed to guard structures, which are grounded microstrips between signal lines that provide limited isolation beyond a few hundred \si{\mega\hertz}~\cite{pajares_ribo_regue_rodriguez-cepeda_pradell}, fences also work at higher frequencies. The shielding effectiveness at a particular frequency \(\nu\) depends on the via spacing. The spacing between vias should remain small compared to the wavelength \(\lambda=v_{\rm m}/\nu\) with the material dependent wave velocity of the waveguide, \(v_{\rm m}\). As a rule of thumb, the spacing should not exceed \(\lambda / 20\) to ensure that the via fence appears solid to an impinging wave~\cite{bahl_2003}, and to minimize loading on the signal-carrying line, which can affect signal integrity~\cite{suntives_khajooeizadeh_abhari_2006}. 
The self-resonance frequency of the vias must also be taken into consideration. The intrinsic shunt inductance of a via can be approximated as $L \approx c_1 h\left(1 + \ln(4h/d)\right)$, with constant \(c_1=\SI{1.95d-6}{\henry\per\meter}\), height $h$ and diameter $d$, and a parasitic capacitance $C \approx c_2 \epsilon_{\rm r} h d_1/(d_1 - d_2)$, where \(c_2=\SI{5.6d-11}{\farad\per\meter}\), $d_1$ is the diameter of the antipad (gap opening in the surrounding ground plane), $d_2$ is the diameter of the via pad~\cite{johnson_graham_1993}, and $\epsilon_{\rm r}$ is the relative permittivity of the dielectric. This leads to a self-resonance for a single via typically in the range of a few hundred \si{\mega\hertz} to a few \si{\giga\hertz}. Furthermore, the combination of the vias' conductivity and the  capacitance formed between the two large ground planes can result in a resonant mode. In both cases, a large number of vias can ameliorate these issues and increase the resonance frequency beyond the qubit frequency operation range.

Using these measures, the crosstalk between directly neighboring control and readout lines is suppressed to below \SI{-40}{\decibel} and next-nearest neighbors to \SI{-60}{\decibel} up to \SI{10}{\giga\hertz}, presented in Fig.~\ref{fig:interposer}.

\subsection{\label{sub:wirebonds}Wirebonding considerations}

We employ thermosonically bonded \SI{25}{\micro\meter} diameter aluminum wirebonds to allow reconfigurable connections and enable rapid prototyping \ReMo{(gold wirebonds can similarly be used as an alternative)}. This also ensures compatibility with 3D-integrated multi-chip modules that use a silicon interposer to fan-out signals to pads for wirebonding~\cite{Rosenberg2017_package}. Alternatively, vertical spring-loaded contacts can be directly fanned out into coaxial cables~\cite{Bejanin2016_package} or connected to a multilayer PCB~\cite{Bronn2018_package}. The device can also be directly clamped onto the PCB~\cite{liu_flip_chip_2017} in a manner similar to flip-chip ball grid array packaging used in conventional room-temperature electronics~\cite{ceramic_bga, flip_bga}, although dielectric losses induced by the proximity of the PCB substrate and surface roughness may reduce qubit coherence~\cite{Rosenberg2017_package}. Inserting a pristine silicon or sapphire interposer between the PCB interposer and qubit chip~\cite{Rosenberg2017_package, Arute2019_supremacy} can reduce such performance limiting effects. These interposer stacks can be constructed using superconducting indium bump-bonds and superconducting through-silicon vias~\cite{tsv_yost_2019}, enabling the construction of multi-chip modules and off-chip resonators~\cite{Rosenberg2017_package}. As in the demonstrated package, a combination of these interconnect techniques can be utilized, such as the use of a silicon interposer with fixed-length wirebonding to the PCB.

The wirebond impedance is dominated by its parasitic inductance. For a wirebond with a diameter of \SI{25}{\micro\meter}, the inductance roughly scales at \SI{1}{\nano\henry} per \SI{1}{\milli\meter} in length~\cite{UCSBWireBond}. Taking into account the effect of a ground plane at a distance \(h\), the inductance of a wirebond with a diameter \(d\) and length \(\ell\) can be approximated by $L \approx \mu_0 \ell \, \rm{arcosh}(2h/d)/(2\pi)$ with the vacuum permeability \(\mu_0\)~\cite{paul_2010}. Similarly, the parasitic capacitance of the wirebond can be approximated with a wire and a uniform metal sheet as $C=2 \pi \ell \epsilon_0/(\rm{arcosh}(h/d))$ with \(\epsilon_0\) being the permittivity of vacuum~\cite{Snow1954_Cap, Iossel1969_Cap}. 

For wirebonds \SI{1}{\milli\meter} in length, the inductance is around \SI{1}{\nano\henry} and the capacitance is in the range of \SI{20}{\femto\farad}. Modeling the wirebond as a transmission line, a \SI{1}{\milli\meter} long wirebond yields a characteristic impedance of \(Z=\sqrt{L/C}=\SI{223}{\ohm}\) and a reflection coefficient of $\Gamma=(Z-Z_{0})/(Z+Z_{0}) \approx 0.63$. Reducing the impedance mismatch requires shorter wirebonds, several wirebonds in parallel, or measures to decrease the parasitic capacitance. 

Mutual coupling between parallel wirebonds hinders the combined inductance from decreasing as quickly as the inverse of the number of wirebonds: about five parallel \SI{1}{\milli\meter} long wirebonds are required to reduce the impedance to the range of \SI{50}{\ohm}. The parasitic inductance can be reduced by positioning signal launches close to the edge of the chip, thus shortening the wirebond, and by using several parallel wirebonds for each signal connection, spreading them in a V-shape to minimize mutual inductance.

In applications which need a low insertion loss, or when the length of the wirebond cannot be minimized, an impedance matching structure can be used~\cite{beer_ripka_2011}. The capacitance is tuned using capacitive structures, such as flares or ``matching dots,'' which are empty metal pads on the interposer that can be galvanically connected to change the capacitance~\cite{whitaker_2005}. 

\begin{figure*}
\includegraphics[width=1\textwidth]{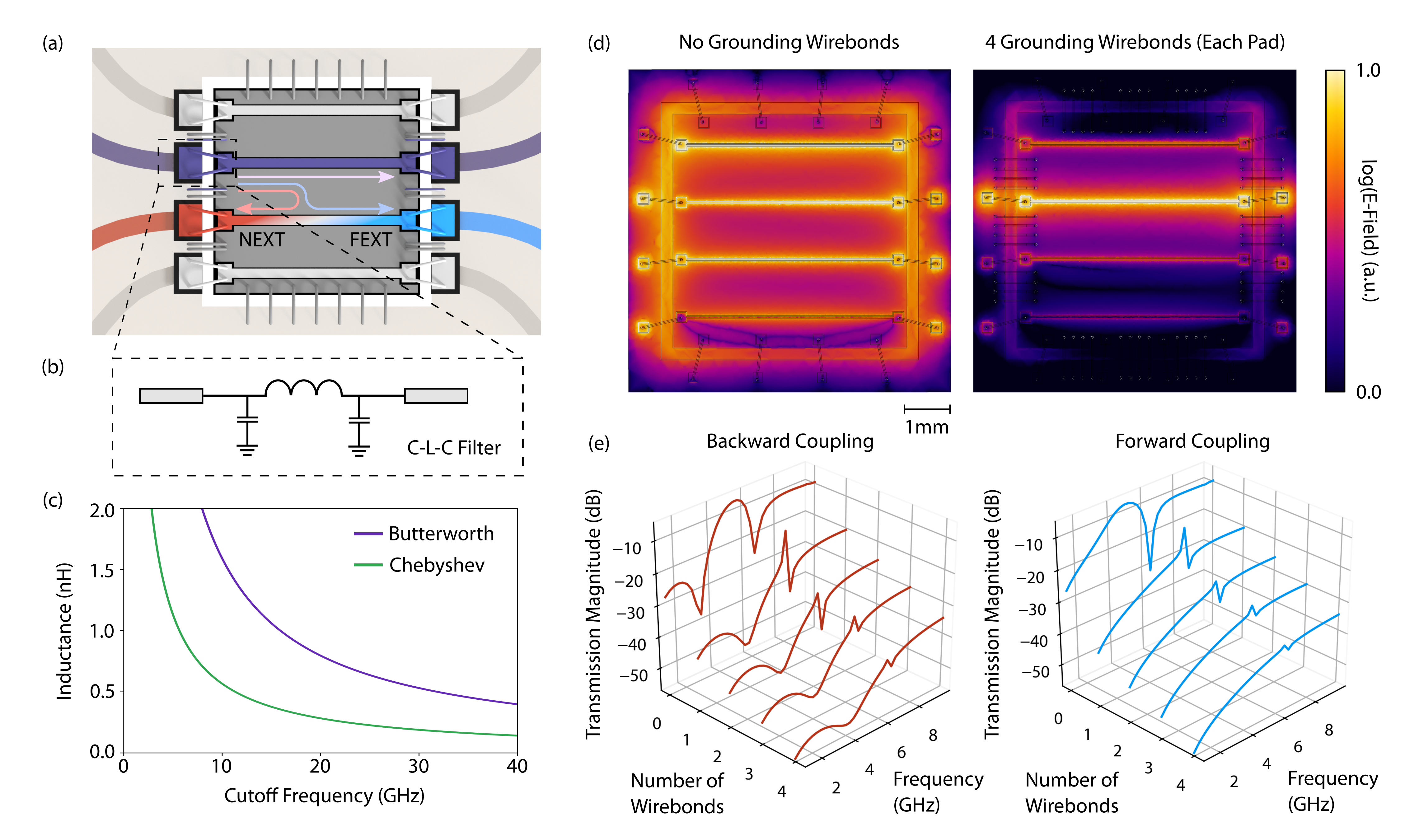}
\caption{\label{fig:wirebond}Wirebond design. (a) Diagram of a chip (gray) with four transmission lines wirebonded to an interposer (beige) surrounding it. The instigating line, indicated in purple, can lead to near-end (NEXT) and far-end (\ReMo{FEXT}) crosstalk in the disturbed line, shaded red and blue respectively. (b) Schematic representation of a wirebond interconnect and its lumped element model. The flares located on the ends of the interposer and chip's transmission lines, as highlighted by the dashed box in (a), correspond to the two tuning capacitors to ground, while the wirebond forms the series inductance. Note that the wirebonds used to connect the signal lines are spread out in a V-shape to minimize mutual inductance. (c) Plot of maximum inductance compensation versus cutoff frequency for Butterworth and Chebyshev filter designs, with \SI{1}{\nano\henry} roughly corresponding to \SI{1}{\milli\meter} in wirebond length. (d) and (e) The effect of wirebond configuration on signal crosstalk. \ReMo{A simplified model with the same spatial dimensions as the 24-pin design is employed to simulate the relative reduction of signal wirebond crosstalk with the addition of more grounding wirebonds.} As the number of grounding wirebonds increases, the electric field \ReMo{strength} between the chip ground and the package cavity rapidly decreases, resulting in the suppressed coupling between adjacent \ReMo{signal wirebonds}.}
\end{figure*}

For basic applications without a bandwidth-limit, an inductor-capacitor structure can be utilized with an impedance of $Z = \sqrt{L_{\rm parasitic}/(C_{\rm parasitic} + C_{\rm tuning})}$. The passband can be increased using a third order, capacitor-inductor-capacitor low pass filter. For a given design cutoff frequency (typically set above the qubit frequency operating range), there is a maximum inductance that can be accommodated in this way due to the broadband requirement. As shown in Fig.~\ref{fig:wirebond}(b), a Butterworth filter can compensate for a wirebond with an inductance \SI{1.6}{\nano\henry} at a cutoff frequency of \SI{10}{\giga\hertz} (corresponding to a length of approximately \SI{1.6}{\milli\meter}), while a Chebyshev configuration can compensate for around \SI{0.6}{\nano\henry}.

The physical distance and the exposure of wirebonds contribute significantly to signal crosstalk, and the combination of the parasitic inductance and capacitance of the wirebond, along with the capacitance formed between the chip and the package ground, form a resonance mode. An impedance ladder model would predict that such crosstalk falls off exponentially with distance at low frequencies, but reaches unity at the resonance frequency of the aforementioned mode~\cite{UCSBWireBond}. As a result, pulling back the ground plane below the chip to decrease the capacitance (for example, designing a cavity underneath the qubit chip) and increasing wirebond density to lessen the inductance serve to reduce signal crosstalk.

EM simulations confirm the formation of a coupling channel due to the presence of a resonance mode between the chip and the chip-mount. This is visualized in Fig.~\ref{fig:wirebond} (c) as a high electric-field density surrounding the chip. Without grounding wirebonds, our particular model, which uses a square \SI{5x5}{\milli\meter} chip, predicts a maximum crosstalk of \SI{-8}{\decibel} that peaks around \SI{5.5}{\giga\hertz}. Increasing the grounding wirebond density drives up the resonance frequency of this mode and expontentially suppresses the crosstalk, with both far- and near-end coupling dropping to below \SI{-40}{\decibel} up to \SI{8}{\giga\hertz} when four wirebonds are employed between signal lines, as shown in Fig.~\ref{fig:wirebond} (d). In the demonstrated package, two wirebonds between each pair of signal lines suppress crosstalk to less than \SI{-30}{\decibel}, at which the performance limitation is negligible relative to other system limitations imposed by on-chip and interposer crosstalk. 

\section{\label{sec:modes}Package Modes}

\begin{figure}
\includegraphics[width=0.47\textwidth]{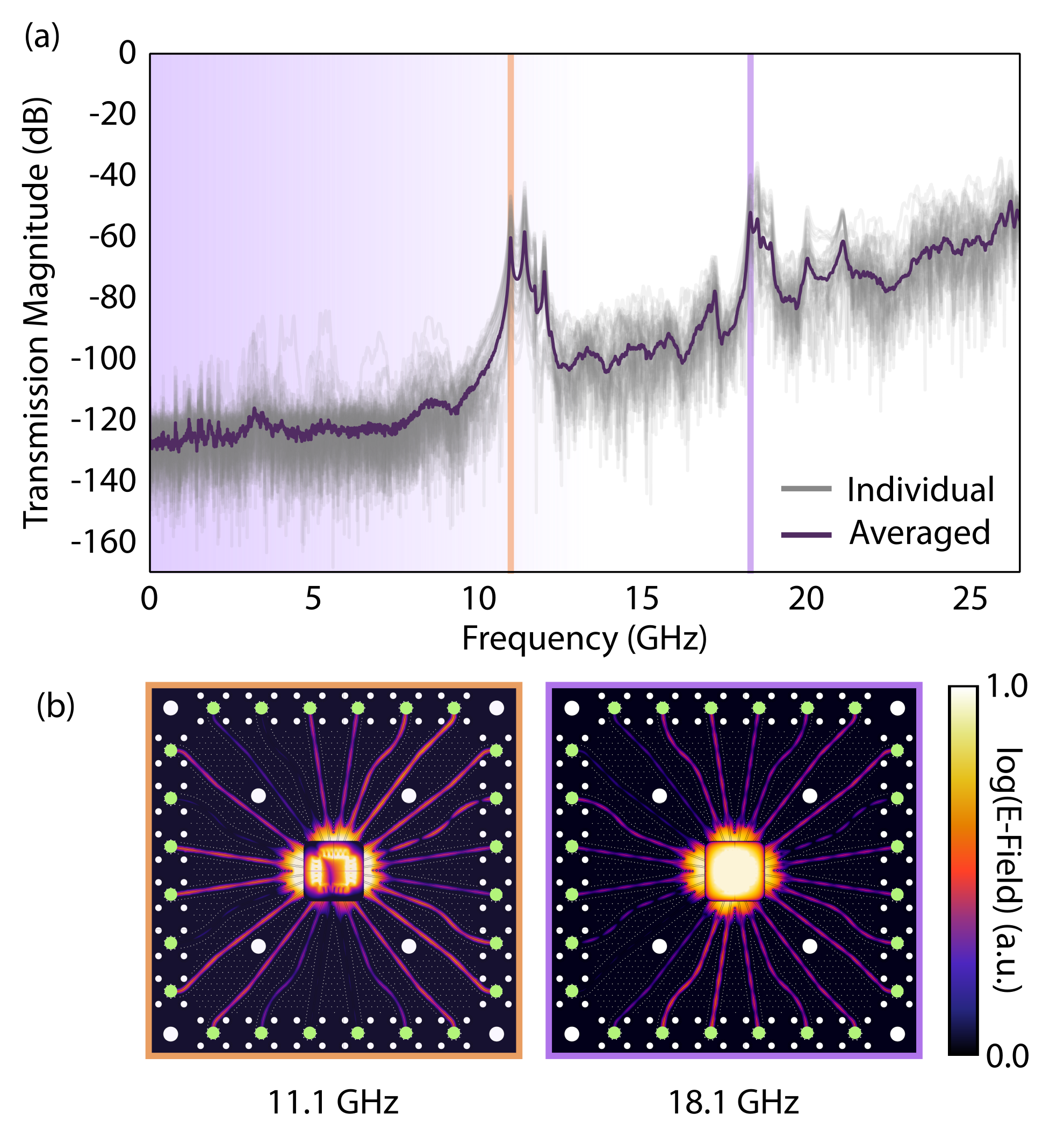}
\caption{\label{fig:simulations}Package mode measurements and simulation. (a) Package modes are probed at liquid nitrogen temperature using a multiport vector network analyzer. The scattering matrix elements corresponding to transmission across the package cavity, as measured via the ports marked green in (b), are overlaid (gray) and averaged (purple), with the relevant frequency range indicated by the background shading. (b) EM simulations reveal eigenmodes at \SI{11.1}{\giga\hertz} (orange) and \SI{18.1}{\giga\hertz} (purple) respectively, which correspond well with the measured peaks as marked by the correspondingly colored vertical lines in (a). The purple background fading out with increasing frequency in (a) indicates the decreasing relevance of modes as their frequency separation to the qubit transition frequency increases.}
\end{figure}

In addition to crosstalk, suppressing package modes is key to a successful design. These resonance modes can reduce the qubit lifetime and induce decoherence. The interaction can be modeled by a two-level quantum system---the qubit---coupled to an EM cavity---the resonant mode---with a rate \(g\). For a small detuning \(\Delta = |\omega_{\rm q} - \omega_{\rm m}| \ll g\) between the qubit angular frequency \(\omega_{\rm q}\) and the package mode angular frequency \(\omega_{\rm m}\), their energy levels hybridize and excitations are coherently swapped between the qubit and the mode. However, since the package modes are often lossy---i.e. have a low quality factor \(Q_{\rm m}\)---they lead to a reduction in the qubit lifetime.

The coupling between a far-detuned mode and a qubit can be described by the Jaynes-Cummings model in the dispersive approximation~\cite{krantz_kjaergaard_yan_orlando_gustavsson_oliver_2019}.

A mode coupling to a qubit ac-Stark shifts the qubit transition frequency by an amount proportional to the average number of photons \(\bar{n}\) present in the mode. Fluctuating photon numbers fluctuate the qubit frequency and induce pure dephasing of the qubit at a rate 

\begin{align}
    \Gamma_{\upphi} = \frac{\kappa}{1+\frac{\kappa^2\Delta^2}{\ReMo{4} g^4}} \bar{n}
\end{align}

\noindent
with \(\kappa=\omega_{\rm m}/Q_{\rm m}\), the decay rate of the mode~\cite{PhysRevLett.94.123602}.

While these wide-band fluctuations within several \si{\giga\hertz} of the qubit transition frequency lead to pure dephasing, modes on the order of \si{\mega\hertz} detuned from the qubit transition frequency can also lead to qubit energy decay due to the Purcell effect, thus reducing the qubit lifetime. If the coupling of the resonant mode to the qubit is small compared to the frequency separation between them, the qubit energy decay rate is

\begin{align}\label{purcell}
\gamma^{\rm Purcell}_{\rm pkg~mode}=\frac{g^2 \kappa}{(\omega_{\rm m} - \omega_{\rm q})^2}.
\end{align}

\noindent
The cavity and interposer geometries often necessary in the design of a package can support resonant modes. To ensure high-fidelity qubit performance, either the coupling of these package modes to the qubit need to be suppressed or their resonance frequency must be far detuned from the qubit operational frequency spectrum. 

\subsection{\label{subsec:box}Box Modes}

The first class of these modes, here referred to as box modes, arise in the enclosing metal cavity and interact directly with the qubit. A box-like cavity is often used to reduce radiative losses from the qubit and significantly cut down on the number of environmental modes at the expense of \(Q\)-enhancement of the modes that remain. Furthermore, the walls of this cavity should be offset from the qubits to reduce material-induced losses, as introduced in Sec.~\ref{sec:mat}, including the "floor" below the chip. As a result, the space above and below the chip forms resonant cavities. The frequencies of the transverse electric and magnetic mode (\(\rm{TE}_{\textit{nml}}\) and \(\rm{TM}_{\textit{nml}}\)) in a rectangular cavity are given as

\begin{align}
    f_{nml} = \frac{c}{2\pi \sqrt{\mu_{\rm r}\epsilon_{\rm r}}} \sqrt{\left(\frac{n\pi}{b}\right)^2+\left(\frac{m\pi}{a}\right)^2+\left(\frac{l\pi}{d}\right)^2}
\end{align}

\noindent
where $\rm{TE}_{\textit{101}}$ and $\rm{TM}_{\textit{110}}$ are the lowest frequency modes, $\mu_{\rm r}$ is the relative permeability, $\epsilon_{\rm r}$ is the relative permittivity, $c$ is the speed of light, and $a$, $b$, $d$ are the three dimensions of the cavity. The components \(\mu_{\rm r}\), \(\epsilon_{\rm r}\), and \(c\) depend on the mode carrying medium, typically the dielectric. The inclusion of any media with an \(\epsilon_{\rm r}\) in excess of the vacuum permittivity, shifts the resonance frequency down. Higher electric field densities around these dielectrics, caused by structures such as wirebonds, can further reduce the mode frequency by increasing the effective dielectric constant~\cite{PozarCavities}. 

\subsection{\label{subsec:slotline}Interposer-Based Modes}

Modes also arise within the package interposer. \textit{Slotlines}, composed of two metallic planes separated by a dielectric gap, support a quasi-transverse electric mode propagating along it. Slotline modes arise when there is a high impedance between two ground planes, for example due to poor galvanic contact between cavity components (such as the lid and the body of the package), and the gap separating the chip and the interposer. Furthermore, the gaps between the signal trace and the ground planes on coplanar waveguides support slotline modes that can be excited by discontinuities or asymmetries~\cite{Slotline}, such as T-junctions and sharp bends on the order of the signal wavelength~\cite{Schuster2001}.

The resonant frequency of these spurious modes can be increased by reducing the impedance between ground planes using wirebonds, vias, or crossovers~\cite{crossover}. Good mechanical contact between cavity or package components can be ensured with tighter machining tolerances or indium crush seals~\cite{stewart2010_indium}.

Similarly, modes can form due to impedance mismatched or open transmission lines with frequencies $f = c/(\ell \sqrt{\epsilon_{\rm r}})$ on the order of a few \si{\giga\hertz} for waveguides with length $\ell$ in the range of centimeters. In the presented package, transmission lines that are not terminated have quarter-wave modes ranging from \SIrange{1.15}{1.38}{\giga\hertz} and can couple to the qubits through unused bond pads. This coupling can be suppressed by grounding unused bond pads using wirebonds. 

\begin{figure*}[t]
\includegraphics[width=1\textwidth]{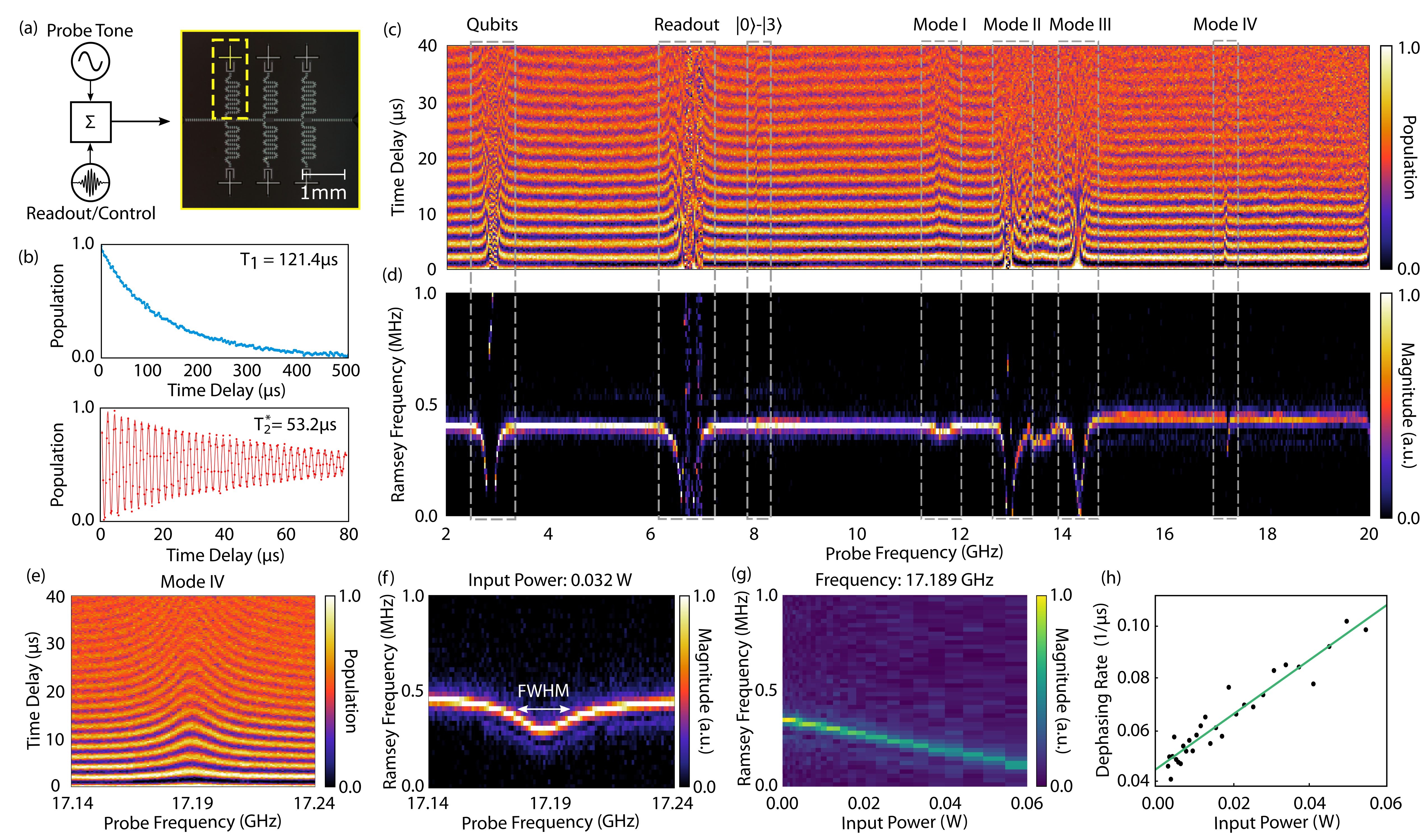}
\caption{\label{fig:HiddenMode}Experimental package mode characterization. (a) Basic setup to measure fixed frequency transmon qubits with individual readout resonators coupling to a common transmission line. In the following panels, the results of the qubit and resonator indicated with a yellow dashed rectangle are explicated. (b) The qubit has an average qubit lifetime of $T_1 \approx \SI{121.4}{\micro\second}$ and coherence time of $T_2^* \approx \SI{53.2}{\micro\second}$ measured in intervals across a period of 12 hours. (c) and (d) The microwave environment between \SI{2}{\giga\hertz} and \SI{20}{\giga\hertz} registered by the qubit is mapped out using the qubit itself as a sensor and a continuous wave probe tone---added to the qubit readout and control signal---interacting with the qubit microwave environment. Probe tone dependent indirect and direct effects on the qubit are recorded using Ramsey spectroscopy. In both the time-domain (c) and frequency-domain (d) panels, the qubits, readout resonators, and the ground to third excited state transition are identifiable as shifts in the Ramsey frequency. Furthermore, four features between \SI{11}{\giga\hertz} and \SI{18}{\giga\hertz} can be identified. The following mode characterization procedure is exemplified on the fourth mode, the expected package cavity mode shown augmented again in (e) at \SI{17.18}{\giga\hertz}. (f) The linewidth of a mode \(\kappa/(2\pi)\) is the full width at half maximum (FWHM) of the Fourier transformed frequency-dependent Ramsey scan, here \SI{20}{\mega\hertz}. (g) The Ramsey frequency change as the power of the probe tone---parked at the mode's resonance frequency, here at \SI{17.189}{\giga\hertz}---is varied. The power is measured at the signal generator. (h) The qubit dephasing as a result of the varying probe tone power is extrapolated by performing fits to $T_2^*$ experiments, as shown in the example in panel (b). The presented analysis yields a coupling rate of \ReMo{$g/2\pi \approx \SI{17.73}{\mega\hertz}$} for Mode IV. Note, for the measurements shown in (c-h) the traveling-wave parametric amplifier (TWPA) was bypassed to prevent interference with the probe tone}
\end{figure*}

\subsection{\label{subsec:substrate}Chip-Based Modes}

Spurious modes also arise from the device substrate, which has a higher $\epsilon_{\rm r}$ than vacuum. For a silicon qubit chip, the first eigenmode ($\rm TM_{\textit{110}}$) is expected at \SI{12.41}{\giga\hertz} for a \SI{5}{\milli\meter}$\times$\SI{5}{\milli\meter} chip, while for dimensions of \SI{10}{\milli\meter}$\times$\SI{10}{\milli\meter}, the first eigenmode drops to \SI{6.20}{\giga\hertz}. These modes can couple very strongly to the qubit due to their physical proximity. One approach to suppress the chip modes is through a change of geometry from a square to a rectangular shape. For example, a \SI{20}{\milli\meter}$\times$\SI{5}{\milli\meter} device, which has the same area as the \SI{10}{\milli\meter}$\times$\SI{10}{\milli\meter} layout, has its lowest eigenmode at \SI{9.04}{\giga\hertz}. Alternatively, through-silicon vias in the qubit chip can be used to pin the resonating modes. Metallization on the chip edge can also be employed to enforce the cavity boundary conditions regardless of the surrounding environment to provide a predictable mode environment.

\subsection{\label{subsec:characterization}Mode Characterization}

A combination of two techniques can be used to identify package modes. First, using a multi-port network analyzer, the average of transmission parameters for geometrically opposing ports---different sides relative to the interposer aperture---result in a transmission spectrum as displayed in Fig.~\ref{fig:simulations}(a). This method relies on the coupling between signal launches in \ReMo{an unpopulated chip cavity (package and interposer without a chip)} and the resonances to resolve spurious modes in the system. A package eigenmode can be resolved in this way because crosstalk across the interposer aperture---typically less than \SI{-100}{\decibel}---is significantly weaker than the transmission induced by the package eigenmode itself. The signal-to-noise ratio is further improved by performing the measurement in liquid nitrogen at $\sim\SI{77}{\kelvin}$, which increases the quality factor of the resonance modes. Measuring across a single pair of connectors does not paint a complete picture, but by taking a large number of scattering parameters, modes around \SI{11}{\giga\hertz} and \SI{18}{\giga\hertz} in our package can be clearly resolved. In our case, we map the mode profile by constructing the full scattering matrix of the package using repeated measurements on a 20-port network analyzer, and averaging the 144 traces that correspond to transmission across the package cavity. The two dominant peaks in the experimental results are in agreement with the 3D EM simulations of the full design, shown in Fig.~\ref{fig:simulations}(b), which revealed two high-\(Q\) eigenmodes at \SI{11.1}{\giga\hertz} and \SI{18.1}{\giga\hertz} respectively.

Detecting modes using a transmission spectrum has two main restrictions. The qubit may couple to resonance modes not visible to signal launches on the cavity periphery, such as chip modes and resonances resulting from device wirebonding. Conversely, the method may detect modes that do not affect qubit operation, such as resonances localized within the interposer. Second, due to the indirect nature of the transmission measurements, the plot in Fig.~\ref{fig:simulations}(a) can only be used as a qualitative tool as the relative amplitudes of the peaks as well as the mode's coupling strength to the qubits cannot be accurately established. 

A second, qubit-based technique for probing package modes is the hidden-mode experiment~\cite{HiddenModeExperiment}, where a fixed-frequency qubit can be used as a mode sensor. It is preferable to use qubits with a long coherence time and stable baseline Ramsey oscillations in order to resolve finer mode structures. A continuous-wave probe tone is injected into the package, either through the readout line or a dedicated port, and swept through the frequency range of interest. For each probe frequency, a $T_{2}$ measurement is performed using Ramsey interferometry on a fixed-frequency qubit. As the probe frequency sweeps in-resonance with a package mode, the package mode will be populated through the coupling between the transmission line and the mode itself. Depending on the mode photon number fluctuations, the coupling to the qubit, and its detuning, the qubit will dephase. The degree of induced dephasing can be inferred with a $T_{2}$ measurement. Due to the wide-band nature of the mode-induced qubit energy level shift, the ac-Stark effect, this technique provides the advantage of facilitating mode measurements across a broad frequency range, typically several tens of \si{\giga\hertz}. 

Fig.~\ref{fig:HiddenMode}(a) depicts a device composed of six superconducting high-coherence transmon qubits used to characterize the package modes. \ReMo{A simple two-port device is chosen to limit the interference of the device with the hidden-mode survey.} Each of these fixed-frequency qubits is weakly coupled to a readout resonator, which is in turn coupled to a shared transmission line. The readout and control signals for all six qubits are frequency-multiplexed and combined with a probe tone from a tunable coherent source, illustrated in Fig.~\ref{fig:HiddenMode}(a). While the probe tone can be injected into the package through control lines, unused signal launches, or a bandpass filter designed to limit qubit energy decays due to the Purcell effect~\cite{Sete2015_Purcell}, we opt to use a transmission line which ideally couples to potential modes homogeneously across the frequency spectrum. Furthermore, as the magnitude of the measured Ramsey frequency shift is roughly proportional to $1/\Delta$, we change the probe power proportional to \(\Delta\)---the frequency difference between the probe tone and the qubit---to resolve modes evenly across the spectrum of interest.

\begin{table}[t]
\center
\setlength{\tabcolsep}{9pt}
\begin{tabular*}{0.45\textwidth}{lcccc}
\toprule
Mode & \(\omega_m/(2\pi)\) &  \(\kappa/(2\pi)\) & \(g/(2\pi)\) & \(T_{\rm pkg~mode}^{\rm Purcell}\) \\ 
& (\si{\giga\hertz}) & (\si{\mega\hertz}) &  (\si{\mega\hertz}) &  (\si{\milli\second}) \\ 
\midrule
I       & 11.65    & 25 & \ReMo{13.05}     & 2.77   \\
II      & 12.94    & 53 & \ReMo{14.15}     & 1.48   \\
III     & 14.30    & 81 & \ReMo{18.23}     & 0.73   \\
IV      & 17.18    & 20 & \ReMo{17.73}     & 5.08   \\ \bottomrule
\end{tabular*}
\caption{\label{tab:modes}The four identified package modes between \SI{2}{\giga\hertz} and \SI{20}{\giga\hertz}. The table contains the mode's resonance frequency \(\omega_m/(2\pi)\), the mode linewidth \(\kappa/(2\pi)\), coupling strength to the qubit \(g/(2\pi)\) as extrapolated using Eq.~\ref{eq:coupling}, and the resulting qubit energy relaxation time \(T_{\rm pkg~mode}^{\rm Purcell}=1/\gamma_{\rm pkg~mode}^{\rm Purcell}\) (Eq.~\ref{purcell}).}
\end{table}

Four package modes are identified between \SI{2}{\giga\hertz} and \SI{20}{\giga\hertz}. The linewidth \(\kappa\) of each spurious mode---a measure of its lossiness---can be determined directly by performing a fine frequency sweep of the hidden-mode experiment. We can thus calculate the strength of the coupling, shown for the fourth mode in Fig.~\ref{fig:HiddenMode} (f). The ac-Stark shift is proportional to the average number of photons \(\Delta_{\rm Stark}=\alpha \bar{n}\) with a factor \(\alpha\). Similarly, the mode induced dephasing is proportional to \(\beta \bar{n}\) with a proportionality factor \(\beta\). The measured $\Gamma_2^*$ is given by 

\begin{align}
    \Gamma_2^* = \Gamma_{\rm mode} + \frac{1}{T_{2, \rm intrinsic}^*} = \beta \bar{n} + \frac{1}{T_{2, \rm intrinsic}^*}.
\end{align}

We can determine \(\alpha\) and confirm the linear relationship by plotting the change in the Ramsey frequency against the input power feeding the mode, and extrapolating the absolute value of the slope, displayed in Fig.~\ref{fig:HiddenMode}(g). The sign of the slope is dependent on whether the qubit drive tone is set above or below the qubit frequency for the measurement. Similarly, \(\beta\) is the slope of the Ramsey decay rate as the input power is increased. Combined, we can calculate the qubit-mode coupling strength

\begin{align}
\label{eq:coupling}
    g = \sqrt{\frac{\beta \kappa (\omega_{\rm m} - \omega_{\rm q})}{4 |\alpha|}},
\end{align}

\noindent

We demonstrate this method by performing a power sweep for Mode IV [Fig.~\ref{fig:HiddenMode}(c)]. Using this method, the coupling strength \(g/2\pi\) was estimated to be \ReMo{\SI{17.73}{\mega\hertz}}. Furthermore, the Purcell limit caused by the mode can be calculated using Eq.~\ref{purcell} to be \SI{5.09}{\milli\second}. Tab.~\ref{tab:modes} summarizes the identified modes and their characteristics. Extrapolating the limit these \ReMo{package modes (pkg mode)} impose on the lifetime of our qubits, \ReMo{using

\begin{align}
\label{eq:purcell}
    T_{\rm pkg~mode}^{\rm Purcell} = 1\Big/\sum_i \gamma_{\rm pkg~mode_i}^{\rm Purcell},
\end{align}
we obtain \(T_{\rm pkg~mode}^{\rm Purcell}=\SI{384}{\micro\second}\) for the modes I-IV.} While this is reasonably sufficient for present devices with coherence times in the range of \SIrange{10}{100}{\micro\second}, we highlight this result because, despite the absence of strong spurious modes up to \SI{11}{\giga\hertz} in our package, a comprehensive survey reveals that the higher frequency modes can still have a significant limiting effect on qubit lifetime. It is likely that these limits will be saturated in the near future as qubit lifetimes increase, underscoring the need for further package design improvements.

This experiment can be further extended to map the spatial distribution of modes within a package. These results are discussed in Appendix~\ref{spatial_hmx}.

\section{\label{sec:con}Conclusion}

Using a newly engineered package, we have validated our approach to package design by systematically examining various elements that can affect superconducting qubit coherence. We present a comprehensive characterization of the effect of package modes on superconducting transmon qubits and corroborate it with results from simulation tools and room-temperature measurements. \ReMo{For our particular qubit design and configuration, the package limits the qubit lifetime to approximately \(T_{\rm limit}^{\rm pkg}=1/(\gamma_{\rm pkg~mode}^{\rm Purcell}+\gamma_{\rm material})=\SI{384}{\micro\second}\). This lifetime is due almost entirely to qubit loss to hidden package modes via the Purcell effect (\(1/\gamma_{\rm pkg~mode}^{\rm Purcell}=\SI{384}{\micro\second}\)). Package material losses (\(1/\gamma_{\rm material}=\SI{9.87}{\second}\)) contribute only at the \SI{15}{\nano\second} level. While the package does not limit the lifetime of the measured qubits, the estimated lifetime limit is within the same order of magnitude of other loss channels.} 
%While the characterization shows that our qubit lifetimes were not limited by the package, the extrapolated \ReMo{material- and package-mode-induced lifetime limit of \(T_{\rm limit}^{\rm pkg}=1/(\gamma_{\rm material}+\gamma_{\rm pkg~mode}^{\rm Purcell})=\SI{384}{\micro\second}\) with \(1/\gamma_{\rm material}=\SI{9.87}{\second}\) and \(1/\gamma_{\rm pkg~mode}^{\rm Purcell}=\SI{384}{\micro\second}\) (a reduction of the package-mode-induced lifetime limit by \SI{15}{\nano\second})} for our particular transmon qubit configuration is within the same order of magnitude of other loss channels. 
Constructing packages with larger devices and qubit lifetimes that are likely achievable in the near future will require a thorough engineering approach that focuses on mode and signal line engineering.

Looking forward, package design will become increasingly critical for larger quantum devices due to their increased complexity. As we increase the number of qubits in today's noisy intermediate-scale quantum (NISQ) devices~\cite{Preskill2018_supremacy}, the precise characterization and suppression of electromagnetic modes and signal crosstalk become even more critical. These established principles for superconducting qubit package are similarly pertinent for future work as packaging techniques are being advanced for systems in the range of 100 to 1000 qubits.

While current state-of-the-art packages still employ wirebonds to provide signal connections between a multi-chip stackup and the device package~\cite{Arute2019_supremacy}, a number of promising candidates such as pogo pins~\cite{Bejanin2016_package}, \ReMo{out-of-plane wiring~\cite{Rahamim2017_PW}}, direct chip-to-interposer connections, and 3D-integrated packaging~\cite{rosenberg_2020} may potentially provide a larger-scale interconnect solution. With greater density, these techniques will face even greater challenges in signal crosstalk, requiring precise impedance matching and the use of shielding structures. The combination of these factors will necessitate thorough simulation and design characterization building on those presented in this manuscript.

\section*{Acknowledgements}

We would like to express our appreciation for Mirabella Pulido for administrative assistance. We thank John Rokosz and Danna Rosenberg for valuable discussions, Joel I. J. Wang for help with device wirebonding, and Keysight Technologies for providing network analysis equipment to perform room-temperature characterization of package modes. This research was funded in part by the ARO grant No. W911NF-18-1-0411; and by the Office of the Director of National Intelligence (ODNI), Intelligence Advanced Research Projects Activity (IARPA) under Air Force Contract No. FA8721-05-C-0002. B.K. gratefully acknowledges support from the National Defense Science and Engineering Graduate Fellowship program. The views and conclusions contained herein are those of the authors and should not be interpreted as necessarily representing the official policies or endorsements, either expressed or implied, of ODNI, IARPA, or the US Government. 

\bibliography{citation}
\clearpage
\newpage
\appendix

\begin{table*}[t]\ReMo{
\begin{tabular}{@{}p{0.1mm} p{3cm} r p{14cm} p{0.1mm}@{}}
\toprule
& Design considerations & & Associated design parameter & \\
\midrule
& \multirow{3}{*}{\shortstack[l]{Material-induced \\qubit-energy losses}} & \(\bullet\) & The distance between the qubit and the package surfaces should be at least \SI{2}{\milli\meter}~\cite{Lienhard2019_package}. & \\
& & \(\bullet\) & In the qubit vicinity (\(\sim\)\SI{2}{\milli\meter}), polymeric materials and silver paste with an electrical conductivity below that of copper should be avoided~\cite{Lienhard2019_package}.& \\ \cmidrule(lr{.75em}){1-5}
& \multirow{4}{*}{Impedance matching}& \(\bullet\) & For connector and waveguide impedances within \SI{10}{\percent} of each other, the reflected power remains below \SI{0.25}{\percent}. & \\
& & \(\bullet\) & For \SI{1}{\milli\meter}-long wirebonds, the capacitance is in the range of \SI{20}{\femto\farad} and the inductance is around \SI{1}{\nano\henry}~\cite{UCSBWireBond} which corresponds to an impedance of \(\SI{223}{\ohm}\). For impedance matching, shorter wirebonds, several wirebonds in a V-shape, or impedance matching structures~\cite{beer_ripka_2011} can be employed. & \\ \cmidrule(lr{.75em}){1-5}
& \multirow{5}{*}{Signal crosstalk} & \(\bullet\) & Symmetrical buried striplines limit signal crosstalk~\cite{PozarCavities,Sage2011_loss}. & \\
& & \(\bullet\) & Via fences reduce crosstalk between waveguides. The spacing between vias should not exceed \(\lambda / 20\) to ensure that the via fence appears solid to an impinging wave at wavelength \(\lambda\)~\cite{bahl_2003}. & \\
& & \(\bullet\) & For a qubit chip mounted with wirebonds, three grounding wirebonds between signal wirebonds reduce the wirebond crosstalk by \(\sim\)\SI{20}{\decibel}. & \\ \cmidrule(lr{.75em}){1-5}
& \multirow{5}{*}{Package-mode profile} & \(\bullet\) & The fundamental package-mode frequency should be greater than twice the maximum qubit transition frequency. The fundamental mode of the package cavity can be increased by breaking the package cavity into subcavities~\cite{Spring2020_Enc}. & \\
& & \(\bullet\) & The chip mode frequency should be greater than twice the maximum qubit transition frequency. The chip mode depends on the chip dimensions and the via positioning~\cite{tsv_yost_2019}. \\\bottomrule
\end{tabular}}
\caption{\label{tab:qualitative}\ReMo{Package design considerations with a focus on qubit-energy-loss channels, impedance matching, signal crosstalk, and package-mode profile.}}
\end{table*}

\section{Interposer design}
\label{appendix:interposer}

\ReMo{The package interposer is built from a laminate composed of two Rogers 4350\texttrademark cores, each with a thickness of \SI{0.338}{\milli\meter}, and bonded using a layer of thermoplastic Fluorinated Ethylene Propylene (FEP) film. The loss tangent of Rogers 4350\texttrademark laminates is relatively low (specified to be 0.0037 at \SI{10}{\giga\hertz} and room temperature). The stackup consists of three copper layers, with signal routing being performed in the center layer and both connector and wirebond launches patterned on the top. To prevent parasitic stub resonances, blind vias are used to route signals between the top and middle layers.

The package is equipped with Rosenberger non-magnetic, gold-plated SMP type connectors. The connectors have a manufacturer-specified insertion loss of $\leq 0.1 \sqrt{f (GHz)} dB$ and offer a frequency range of DC-\SI{40}{\giga\hertz}, covering the operating range of most superconducting qubits.}

\ReMo{The connector transitions are characterized using time-domain reflectometry (TDR) on a \SI{14}{\giga\hertz} bandwidth Keysight E5063 network analyzer. The connector launch is well matched to the waveguide, as shown in Fig.~\ref{sup_fig:TDR}. The interposer waveguide has a measured impedance of \SI{53.5}{\ohm}, leading to a voltage standing wave ratio of $1.07$ and a mismatch loss of \SI{0.05}{\decibel} (\SI{0.1}{\percent} of the incoming power is reflected).}

\begin{figure}[h]
\includegraphics[width=0.45\textwidth]{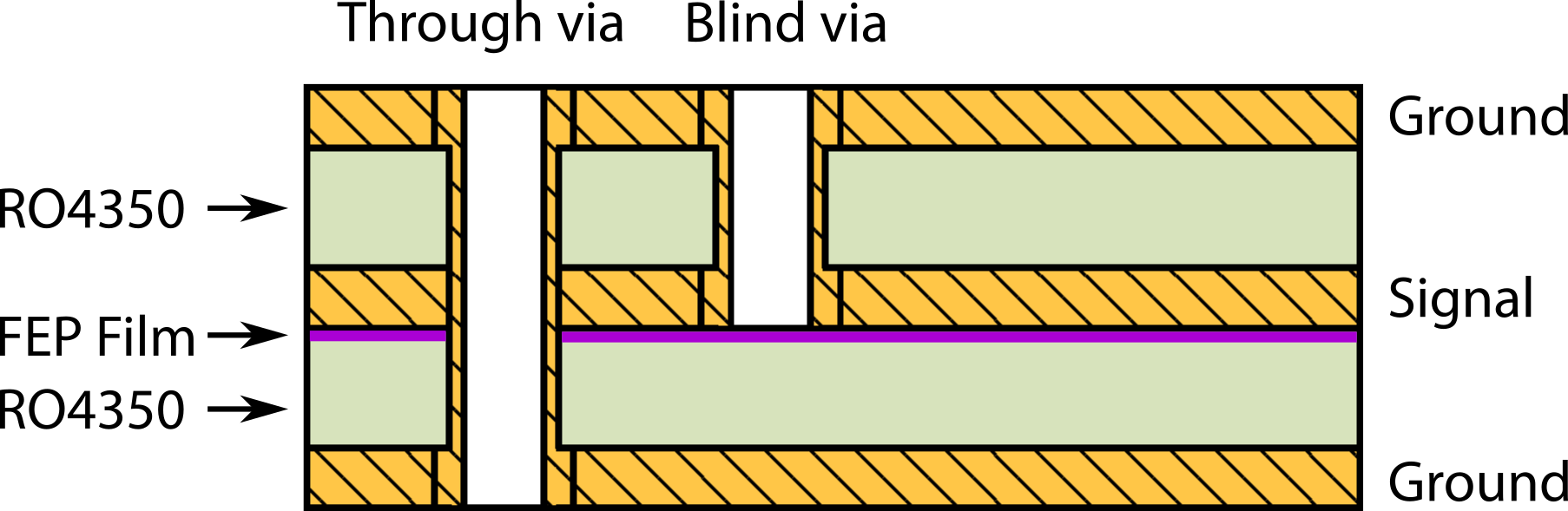}
\caption{\label{sup_fig:stackup} \ReMo{Schematic diagram of the interposer stackup configuration (not to scale). The board consists of two Rogers 4350\texttrademark cores bonded by a layer of FEP film, indicated in purple. Through vias are used for grounding and shielding whereas blind vias are utilized for signal routing to minimize parasitic resonances.}}
\end{figure}

\begin{figure}[h]
\includegraphics[width=0.48\textwidth]{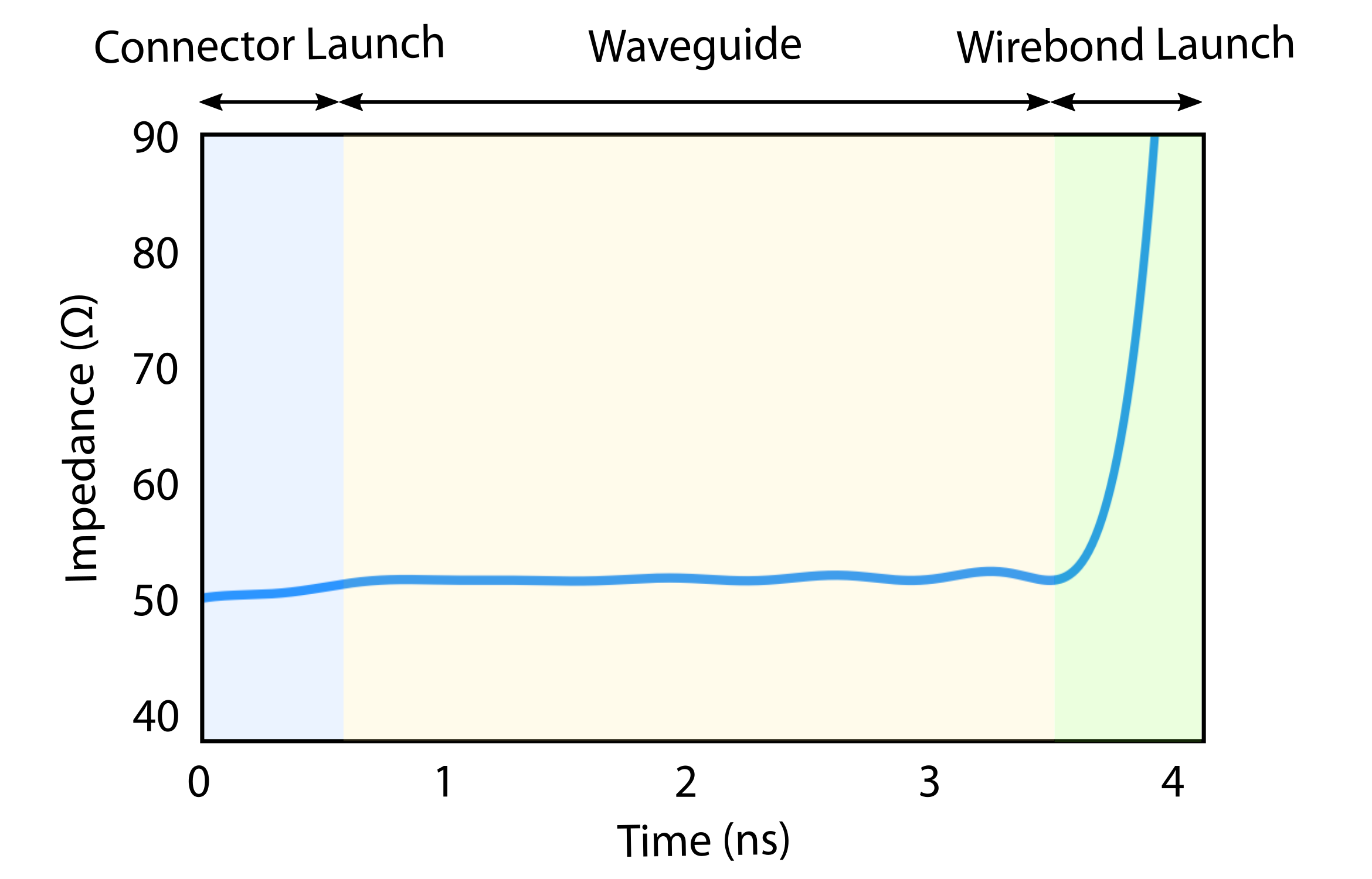}
\caption{\label{sup_fig:TDR} \ReMo{Single-ended TDR measurement of the SMP-to-interposer connector transition. The reference plane of the connector is located at \SI{0}{\nano\second}. The wirebond launch is left open (no chip connected) resulting in a steep increase in impedance.}}
\end{figure}

\section{Measurement setup}

\begin{figure}
\includegraphics[width=0.5\textwidth]{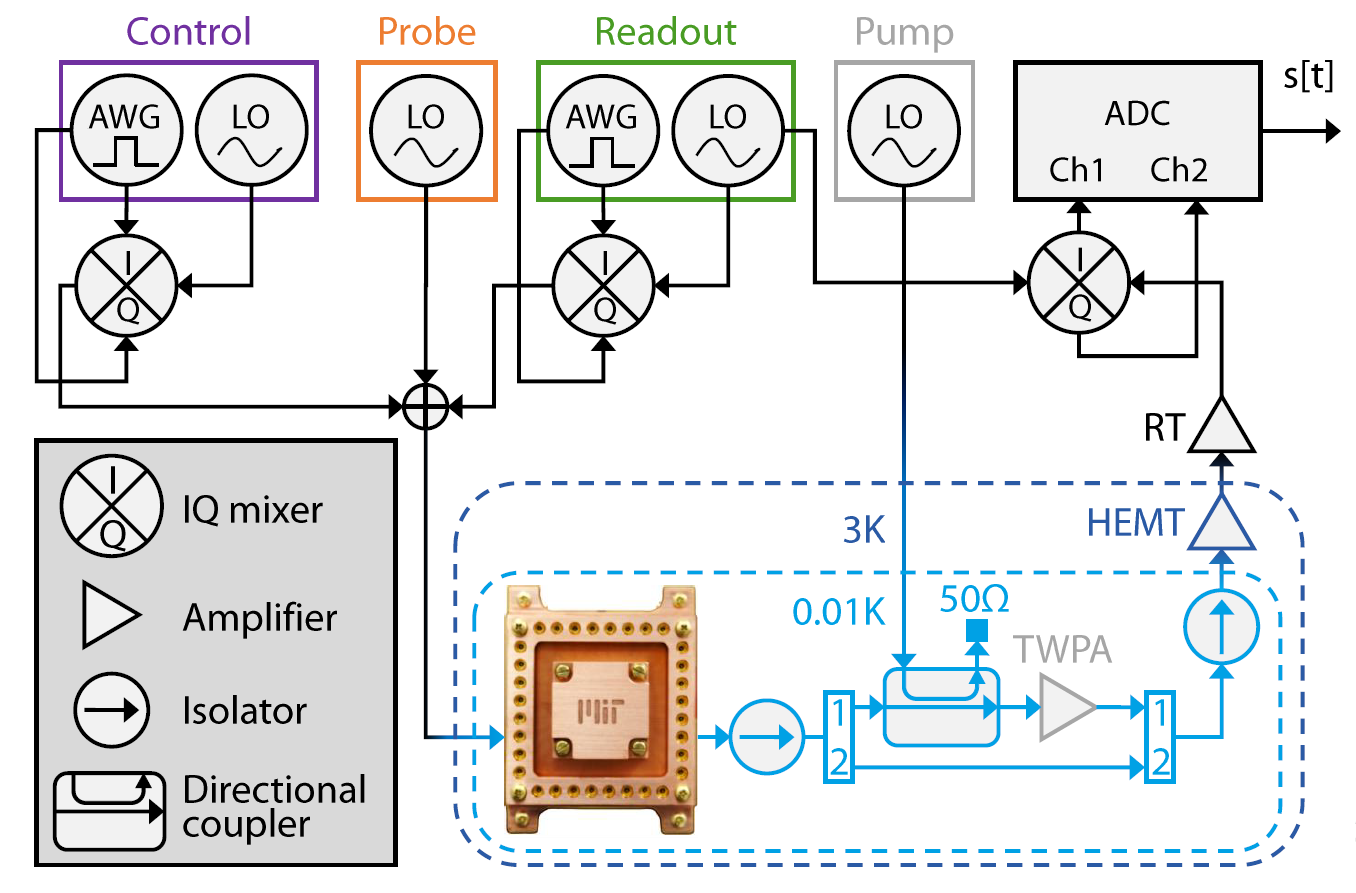}
\caption{\label{sup_fig:fridge} Measurement setup used to obtain the mode profile of the package. The control, readout, and probe signals are combined and sent down the dilution refrigerator with a total of \SI{60}{\decibel} of attenuation distributed at various stages (not shown). The readout signal is amplified by a traveling-wave parametric amplifier (TWPA) at the \SI{10}{\milli\kelvin} stage, a high-electron-mobility transistor (HEMT) amplifier at the \SI{3}{\kelvin} stage, and a low noise amplifier at room temperature before being down-converted and subsequently digitized. Note, the TWPA can be bypassed if necessary.}
\end{figure}

The measurement setup used to perform the hidden mode experiment is depicted in Fig.~\ref{sup_fig:fridge}. Control pulses for the qubits are created using two separate Keysight M3202A PXI arbitrary waveform generators with sampling rates of \ReMo{\SI{1}{\giga\sample/\second}}. The in-phase and quadrature signals are upconverted to the qubit transition frequency using an IQ-mixer which acts as a single side band mixer. The probe tone is created using a separate signal generator. The control, readout, and probe tones are then combined and sent to the dilution refrigerator via a single microwave line.

There is a total of \SI{60}{\decibel} of attenuation distributed within the dilution refrigerator wiring to reduce thermal noise from room temperature and the higher temperature stages of the refrigerator. To reach the qubit, the signal has to pass through the readout resonator, which acts as a filter. A control pulse length ranging from \SIrange{100}{150}{\nano\second} is used to excite the various qubits.

The state of the qubit is determined via dispersive readout. The frequency of the resonators coupled to each qubit change slightly depending on the qubit state. This difference can be measured by sending a measurement tone near the corresponding resonator frequency down the central transmission line and recording the transmitted signal. 

The signal can be first boosted using a traveling-wave parametric amplifier (TWPA) which has a gain of up to \SI{30}{\decibel}, a very low noise temperature, and a wide bandwidth that enables multiplexed readout~\cite{Macklin2015_Science_JTWPA}. The TWPA requires a pump tone, which is sourced from a signal generator at room temperature. The microwave line carrying the pump tone is attenuated by \SI{50}{\decibel} and fed into the TWPA via a set of a directional coupler and isolator located at the \SI{10}{\milli\kelvin} stage of the refrigerator. The signal is further amplified by a Low Noise Factory high-electron-mobility transistor (HEMT) amplifier thermally anchored to the \SI{3}{\kelvin} stage.

At room temperature, the readout signal is fed into a heterodyne detector. The down-converted in-phase and quadrature signals are digitized with a Keysight M3102A PXI Analog to Digital Converter with a \ReMo{\SI{500}{\mega\sample/\second}} sampling rate. The signal is integrated into the internal field-programmable gate array (FPGA) of the digitizer to extract the occupation probability of the qubit in a given state.

\section{Chip design}

The device reported in the main text have a geometry of \SI{5}{\milli\meter} by \SI{5}{\milli\meter} and was fabricated following a similar process as described in~\cite{Yan2016_NatCom_flux} on a high resistivity \SI{275}{\micro\meter} (001) Si wafer ($>$\SI{3500}{\ohm\centi\meter}). The chip consists of aluminum superconducting coplanar waveguides (CPW) and 6 superconducting fixed-frequency transmon qubits around \SI{3}{\giga\hertz}. The qubits are capacitively coupled to individual quarter-wave resonators that couple again inductively to a \SI{50}{\ohm} feedline in the chip's center. The resonator lengths are varied to frequency multiplex the resonances in the range of \SIrange{6.69}{6.81}{\giga\hertz} with a spacing of approximately \SI{25}{\mega\hertz}. 

\begin{table}[b]
\begin{tabular}{@{}ccc@{}}
\toprule
Qubit number~~&~~Qubit frequency~~&~~Readout frequency \\ 
 & (\si{\giga\hertz}) & (\si{\giga\hertz}) \\\midrule
1            & 2.8883                & 6.6874                  \\
2            & 2.9531                & 6.7094                  \\
3            & 3.0527                & 6.7367                  \\
4            & 3.1600                & 6.7599                  \\
5            & 3.2402                & 6.7805                  \\
6            & 3.2572                & 6.8066                  \\ \bottomrule
\end{tabular}
\caption{\label{tab:chip}Qubit chip comprising six fixed-frequency transmon superconducting qubits between \SIrange{2.88}{3.26}{\giga\hertz} and individual readout resonators between \SI{6.69}{\giga\hertz} and \SI{6.81}{\giga\hertz}. The individual readout resonators couple to the common transmission line. On one side of the transmission line are qubits 1, 3, and 5 and on the other side qubit 2, 4, and 6.}
\end{table}

\begin{figure*}[t]
\includegraphics[width=1\textwidth]{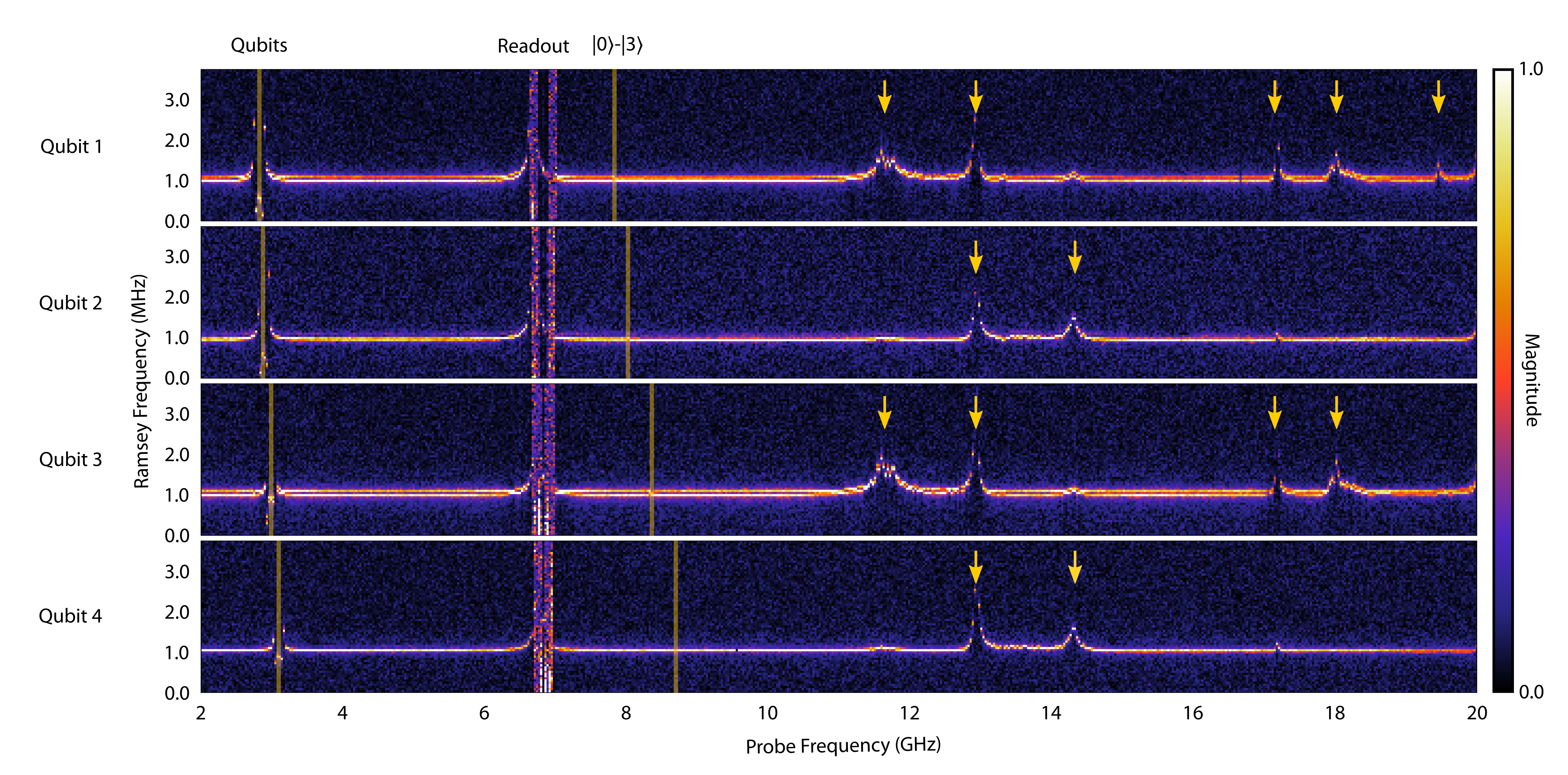}
\caption{\label{sup_fig:Spatial_HMX}Experimental package mode evaluation using four qubits. The Ramsey spectra are taken simultaneously from four different qubits while a probe tone is injected through the central transmission line. \ReMo{Note the similarity between the spurious modes (indicated by yellow arrows) measured by qubits 1 and 3, as well as qubits 2 and 4.} The orange lines indicate qubit-related features which shift depending on the sensor qubit used.}
\end{figure*}

\section{Spatial mode characterization}
\label{spatial_hmx}

An extension of the mode characterization performed in the main text is the use of a multi-qubit device to resolve spatially-dependent modes. Different resonances in the package may couple to the qubits with different strengths depending on their electric field distribution. Such a technique is useful in identifying spurious modes that may be hidden from any given sensor qubit especially when the device area and field variability increases.

We selected qubit 2 for Fig.~\ref{fig:HiddenMode} as it had the longest coherence time. This provides the best resolution in performing the Ramsey sweep. Here, we extend the experiment by performing simultaneous readout on four qubits \ReMo{(the remaining two qubits were not operational due to issues unrelated to the package)}. This came at the cost of a beating effect in the Ramsey oscillations of qubit 1 and 3, as can be seen via the two lines persistent throughout the probe frequency sweep. This was caused by spurious tones arising from instrumentation in our particular multiplexed readout configuration. However, the relevant frequency shifts can still be clearly identified.

\ReMo{The multi-qubit hidden mode survey in Fig. 8 reveals several spurious modes positioned at the same frequency across all qubits. For example, the mode at \SI{11.65}{\giga\hertz} more strongly affects qubits 1 and 3, while the mode at \SI{14.3}{\giga\hertz} interferes with qubits 2 and 4, suggesting spatially-dependent coupling strengths. The mode at \SI{12.94}{\giga\hertz} and the fundamental package mode at \SI{17.18}{\giga\hertz} couple to all qubits and do not reveal a spatially-dependent coupling strength. Finally, several features between \SIrange{2}{9}{\giga\hertz} (highlighted with orange vertical lines) vary in frequency depending on the sensor qubit used. These are qubit-dependent; namely, they are associated with the qubit $\ket{0}$ to $\ket{1}$ and $\ket{0}$ to $\ket{3}$ transitions, as well as the respective readout resonators located around \SI{7}{\giga\hertz}.}

We attribute this to the chip layout: each set are located on one side of the device, separated by a transmission line. The different coupling strengths are likely due to asymmetries in chip placement. For example, an offset that causes stronger capacitive coupling between the chip and the interposer aperture on one side will affect the two sets of qubits differently.

This measurement only revealed two distinct mode profiles due to the symmetric arrangement of the chip. However, a measurement of this type may be useful in probing the mode structure of more complex devices. In larger chips, spurious modes can arise locally within the chip substrate, and qubits at different locations within the package may couple to these modes with varying strengths. Understanding the spatial distribution of modes is needed to ensure the consistent performance of all qubits on the device.

\end{document}